# Carrier localization and out of plane anisotropic magnetoresistance in $Nd_{0.55-x}Sm_xSr_{0.45}MnO_3$ thin films

M. K. Srivastava,[1,2] A. Kaur[2] and H. K. Singh[1#]

[1]*National Physical Laboratory (Council of Scientific and Industrial Research), Dr. K. S. Krishnan Marg, New Delhi-110012, India*
[2]*Department of Physics and Astrophysics, University of Delhi, Delhi-110007, India*

Abstract

The impact of carrier localization on the anisotropic magnetoresistance (AMR) has been investigated in $Nd_{0.55-x}Sm_xSr_{0.45}MnO_3$ (x=0.00 – 0.45) thin films. Carrier localization is caused by the reduced average radius of the A-site of the perovskite lattice and enhanced size disorder due to substitution of smaller $Sm^{3+}$ cations for larger $Nd^{3+}$. This results in (i) enhanced Jahn-Teller (J-T) distortion as evidenced by enhancement in the activation energy of small polaron hopping, (ii) decrease in the ferromagnetic (FM) and insulator-metal transition (IMT) temperatures ($T_C/T_{IM}$), (iii) sharpening of the IMT, (iv) increase in the low field MR, and (v) large enhancement in the AMR. The AMR-T curves of all films show a maximum just below $T_{IM}$. The peak AMR measured at small magnetic field of 4.2 kOe increases from -5 % for x=0.00 to -60 % for x=0.45. The AMR enhancement has been explained in terms of the unquenching of the orbital angular momentum of $t_{2g}^3$ & $e_g^1$ configuration and spin fluctuations due to stronger J-T distortion at higher values of x.

[#]**Corresponding Author**
**Email: hks65@nplindia.org**

Perovskite manganites with a general formula $RE_{1-x}AE_xMnO_3$ where RE is rare earth and AE is alkaline earth have attracted attention because of the occurrence of variety of phenomena, such as paramagnetic insulator (PMI) to ferromagnetic metal (FMM) transition, which is accompanied by the colossal magnetoresistance (CMR), and charge/orbital order (CO/OO) that generally has antiferromagnetic (AFM) spin order.[1-4] Apart from colossal magnetoresistance (CMR), another important property of doped rare earth manganites is anisotropic magneto resistance (AMR) due to anisotropic magneto-crystalline nature that results in the dependence of resistivity on the angle between the applied magnetic field (**H**) and the direction of magnetic easy axis or the transport current (**J**).[5-15] The AMR in manganites has been found to be different from that in FM metals, wherein it shows a peak in the vicinity of the insulator metal transition (IMT). The AMR in manganite thin film has also been found to be sensitive to the nature of the strain. It has been shown by O'Donnel that in compressively strained $Pr_{0.7}Sr_{0.3}MnO_3$ ultrathin films on $LaAlO_3$ substrates, the strain-induced anisotropy field, ($H_K$) which favors an out-of-plane magnetization, is larger than the demagnetization field ($H_D$), resulting in an out-of-plane easy axis.[10,15] In absence of any strain, the threshold filed $H_t$ (=$H_K$ + $H_D$) is small because the anisotropic field $H_K$ partially cancels out $H_D$ and the spontaneous magnetization lies in the plane of the film. It has also been proposed that the magnetotransport anisotropies in manganites could also arise due to a local, spin–orbit induced, orbital deformation that influences the local hopping conduction process characteristic of these strongly correlated oxides in the vicinity of the paramagnetic-ferromagnetic (PM-FM)/IM transition temperature ($T_{IM}$). Li et al.[16] have proposed that the atomic orbital ordering caused by the strain-induced static Jahn–Teller (J-T) distortion and the spin-orbit (SO) coupling play crucial roles in the AMR properties and it depends on the dissimilar response of J-T distortions to external magnetic field. Thus manganites having smaller bandwidth (BW) and hence stronger J-T distortions are expected to have greater AMR effect.

In low bandwidth manganites the enhanced J-T distortion increases the phase competition between the FM and charge/orbital ordered (CO/OO) AFM phases and induces bicritical features. However, Rodriguez-Martinez and Attfield[17] have demonstrated that the competition between the FM and AFM-CO/OO phases is not determined by the 3d electron transfer interaction or the bandwidth alone but the average ionic radius and the size mismatch of the RE and AE ions of the A-site in the perovskite lattice are also important. The latter effect representing the local lattice distortion transmitted from the randomly substituted A-site is measured by the variance, $\sigma^2 = \sum(x_i r_i^2 - <r_A>^2)$, where $x_i$ and $r_i$ are the fractional occupancies and the effective ionic radii of RE and AE cations, respectively. $<r_A>$ is the average ionic radius of the A-site. The local distortion arising from the difference in ionic radii, that is, $\sigma^2$ and/or the random Coulomb potential due to the trivalent/divalent ion mixture is the source of the quenched disorder, which is believed to play a crucial role in determining the magnetotransport in low BW manganites.[1,2,5] Thus in low BW manganites because of the random potential affecting the FM versus CO/OO bicritical feature, magnetotransport properties are appreciably affected.[1] It has been suggested that manganites that have strong quenched disorder would have higher AMR.[5] Here we have systematically studied the impact of BW and size disorder on the AMR in oriented $Nd_{0.55-x}Sm_xSr_{0.45}MnO_3$ thin films.

Ultrasonic nebulized spray pyrolysis[18] was used to prepare polycrystalline $Nd_{0.55-x}Sm_xSr_{0.45}MnO_3$ (x=0.00, 0.05, 0.15, 0.25, 0.35, and 0.45) thin films on single crystal LAO (001) substrates. Film deposition was done at T~200 °C. All the films were annealed in air at 1000 °C for duration of 12 hrs followed by slow cooling. The structural characterization was performed by X-ray diffraction (XRD). The cationic composition was studied by energy dispersive spectroscopy (EDS) attached to scanning electron microscope. The temperature and magnetic field dependent magnetization was measured by a commercial magnetic property measurement system (MPMS, Quantum Design). The electrical resistivity was measured by the standard four probe technique, wherein the current (**J**) was applied in the plane of the film along the longest direction, i.e., it coincided with easy axis. With the direction of **J** fixed in the easy plane, the magnetic field direction was varies from θ=0° (**H**//magnetic easy axis (and also **J**), **H**//film plane) to θ=360° through θ=90° (**H**⊥easy axis plane (and also **J**), **H**//magnetic hard axis) by rotating the film. The AMR measured in the present study is defined as AMR = ($\rho^{in}_{\parallel}$ - $\rho^{out}_{\perp}$)*100/$\rho_{av}$. $\rho^{in}_{\parallel}$ is the resistivity of the film for **H** (maximum **H** applied was 4.2 kOe) applied parallel to the plane of the film. $\rho^{out}_{\perp}$ represents the resistivity measured when **H** is applied perpendicular to the plane of the film. The average resistivity is defined to be $\rho_{av} = \rho^{in}_{\parallel}/3 + 2\rho^{out}_{\perp}/3$.

The films grown by chemical spray pyrolysis are in general found to be polycrystalline but oriented.[18] The thickness was measured to be ~300 nm. The XRD data (θ - 2θ scans) shows the diffraction peaks corresponding to the out of plane lattice constant (c-parameter) of the films appear alongside that of the substrate; suggesting that the films are highly oriented. The representative XRD patterns are shown in Fig. 1. The pristine film (x=0.00) has the out of plane c-parameter of 3.825 Å, which decreases almost linearly with Sm doping to c=3.816 Å for x=0.45. Since the ionic radius of $Sm^{3+}$ (1.24 Å) is slightly smaller than that of $Nd^{3+}$ (1.27 Å), observed decrease in the lattice constant is consistent with increasing Sm content (x).

The FM-PM transition temperature (data not shown here) was observed at $T_C \approx 275$ K in $Nd_{0.55}Sr_{0.45}MnO_3$ (x=0.00) and gradually decreased to $T_C \approx 155$ K for x=0.45. Concomitant to the decrease in $T_C$, the bifurcation between the ZFC and FC M-T curves increases. This suggests the formation of spin cluster glass (CG) state at increased Sm content due to coexisting AFM-COI clusters in the FMM matrix.[18,19] The isothermal M-H loops measured with **H** applied parallel and normal to the film surface showed that in all films the easy axis was in the film plane, while the hard axis was along the normal. In absence of substrate strain, $H_t$ is small and hence the shape anisotropy dominates resulting in the inplane magnetic easy axis.

The temperature dependence of resistivity (ρ-T) is shown in Fig. 2. The pristine (x=0.00) film shows IMT at $T_{IM} \approx 279$ K and this value is in excellent agreement with the phase diagram of $Nd_{1-x}Sr_xMnO_3$.[1] As x is increased, (i) ρ-T in the PM regime shows a sharper rise on decreasing temperature, (ii) the $T_{IM}$ decreases, and (iii) the IMT becomes sharper. The $T_{IM}$ values for x=0.05, 0.15, 0.25, 0.35 and 0.45 are found to be 268 K, 237 K, 218 K, 195 K and 155 K, respectively. The ratio of the resistivity at IMT to that at the room temperature ($\rho_{IMT}/\rho_R$) increases, first slowly from 1.042 for x=0.00 to 5.116 for x=0.35 and then suddenly jumps to 32.654 at x=0.45. The sharpening of the IMT is also well evidenced by the increase in the TCR (temperature coefficient of resistance) ≈ 2 % for x=0.00 to 42 % for =0.45. These variations are understood in terms of the substitution of $Nd^{3+}$ by smaller $Sm^{3+}$ in $Nd_{0.55}Sr_{0.45}MnO_3$ which is very close to the regime of coexisting FMM and AFM- COI phases.[1,19] The doping with $Sm^{3+}$ ions decreases the average RE-site cationic radius ($<r_A>$) and enhances the quenched disorder (as measured by $\sigma^2$), which in turn are expected to (i) enhance the electron-lattice coupling due to the enhanced J-T distortion of the $MnO_6$ octahedra, and (ii) reduce the carrier bandwidth.[1-3,17] Thus the net impact of reduced $<r_A>$ and enhanced quenched disorder (higher $\sigma^2$) is stronger carrier localization.[1-5,17] Variation of $T_{IM}$ as a function of $<r_A>/\sigma^2$ is plotted in the inset of Fig. 2. The observed strong increase in the resistivity just below $T_{IM}$ in the PMI regime, in x=0.35 and 0.45 films is a consequence of carrier localization and also could be due to the enhanced AFM-COI phase.

The variation of temperature dependence of AMR with the Sm content (x) shows a very interesting trend (Fig. 3). In all films the AMR-T curve shows a peak just below $T_{IM}$. In the pristine $Nd_{0.55}Sr_{0.45}MnO_3$ film (x=0.00) the peak AMR ($AMR^{peak}$) is very small (≈ -5%). On partial substitution of Sm for Nd the $AMR^{peak}$ first increases slowly to ≈ -9 % at x=0.15 and then shows steep rise. Thus $AMR^{peak}$ increases from ≈ - 15 % in x=0.25 to ≈ - 59 % at x=0.45. Thus it is clear that the AMR increases with decrease (increase) in $<r_A>$ ($\sigma^2$). The variation of the $AMR^{peak}$ with $<r_A>$ and $\sigma^2$ shown in the inset of Fig. 3 brings out that these two parameters could play crucial role in the enhancement of AMR. Since the two consequence of Sm substitution, viz., the decrease in $<r_A>$ and increase in $\sigma^2$ lower the one electron BW and increase the J-T distortion of the $MnO_6$ octahedra. The enhanced J-T distortion is known to favour the AFM-superexchange (SE) and carrier localization.[1] Thus carrier localization caused by the decrease in $<r_A>$ and enhanced quenched disorder appears to play a crucial role in determining the AMR.

In order to see the correlation between AMR and carrier localization caused by the effect of enhanced J-T distortion of the $MnO_6$ octahedra we analyzed the resistivity in the PMI regime (T > $T_{IM}$) for all the films in the frame work of the Emil-Holstein small polaron hopping model[18,20] given by $\rho(T) = AT\exp(E_A/k_BT)$ where the constant A is related to 'n' the polaron concentration; 'a' the site-to-site hopping distance, and 'ν' the attempt frequency, $k_B$ is the Boltzmann constant, T is the temperature and the $E_A$ is activation energy, which is the height of the potential barrier approximated by $E_A \approx E_P/2$, the polaron binding energy. From the fitted data (not shown here) we calculated that the average activation energy ($E_A$) for each film. The calculated values of $E_A$ are 41.35 meV, 50.76 meV, 71.64 meV, 90.25 meV, 105.49 meV and 120.45 meV, respectively for x=0.00, 0.05, 0.15, 0.25, 0.35 and 0.45 respectively. Enhancement in the activation energy is evidence of the stronger carrier localization due to J-T distortion caused by decrease in $<r_A>$ and size disorder as explained earlier. The dependence of $T_{IM}$ and the $AMR^{peak}$ on the activation energy is plotted in Fig. 4. The $T_{IM}$ is seen to decrease nearly linearly with increasing $E_A$ till x=0.25 and then the drop is sharper. The decrease in $T_{IM}$ is caused by the enhanced carrier localization. The near linear increase in AMR till x=0.15 ($E_A$ = 71.64 meV) is followed by a strong rise in it at higher values of x. The gradual increase in the J-T distortion with increasing Sm content could also enhance its asymmetry. The anisotropic J-T distortions are expected to alter the spin-orbit interaction of the system via changing the magnetic interaction through the manganese ions. It has been suggested that the large AMR in the vicinity of IMT may be the result of the distinct lattice response to external magnetic field because of the anisotropic JT distortions.[16] The increase in AMR with Sm content, which increases the quenched disorder and enhances the J-T distortion of the $MnO_6$ octahedra appears to validate the model of Li et al. wherein it has been proposed that the large AMR in the vicinity of IMT may be the result of the distinct lattice response to external magnetic field because of the asymmetric J-T distortions.[16] The quenched disorder also induces phase fluctuation and enhanced phase competition. The competing phases in the presented case are expected to be FM-metal and AFM-CO-insulator. This in turn enhances the spin fluctuations at T~ $T_C$ and pushes the system towards bicriticality. Application of even small field H ≈ 4.2 kOe suppresses the phase fluctuation and causes the huge low field MR~ 75 % when H // film plane. This is due to the transformation of liquid like AFM-COI clusters into FMM. The small low field MR when H ⊥ film plane suggests that in this configuration the response of the lattice to the magnetic field may be different from the case when H is parallel to the film plane. This indirectly suggests that asymmetry in the J-T distortions could also increase with the Sm content and the quenched disorder. It has been suggested that the large AMR in the vicinity of IMT may be the result of the distinct lattice response to external magnetic field because of the anisotropic JT distortions.[16]

The results described above show that the carrier localization that is determined by the interplay between the two competing interactions, viz., the FM-DE and the J-T distortion of the $MnO_6$ octahedra that favour AFM phase, play the most crucial role in fixing the AMR. The dominance of FM-DE over the J-T distortion in the pristine $Nd_{0.55}Sr_{0.45}MnO_3$ (x=0.00) film, which has the smallest AMR is well evidenced by the smallest $E_A$ (41.35 meV) and the highest $T_C/T_{IM}$. From our results it is clear that increasing Sm content results in strong degree of quenched disorder, which is in turn, is seen to favour reduction in $T_C/T_{IM}$, and strengthens carrier localization due to the growth of AFM/COI phase. This is well evidenced by the increase in the value of $E_A$ with increasing Sm content and hence the variance $\sigma^2$. As the Sm content (x) is increased, owing to its smaller ionic size (i) the J-T distortion increases, as evidenced by the increase in $E_A$, and (ii) the FM-DE is weakened as shown by the decrease in the $T_C/T_{IM}$. Further, when the J-T distortion is smaller, e.g., at x< 0.15, the orbital angular momentum of $t_{2g}^3$ & $e_g^1$ configuration is nearly quenched by the crystal field and hence the spin orbit (SO) coupling is very weak. At higher value of x, the enhanced cooperative J-T distortion is expected to favour the growth of AFM-COI state at the cost of FMM, that is, the competition between FM-DE and AFM-SE is expected to increase. This leads to unquenching of the 3d orbits and the associated angular momentum, hence causing the appearance of SO coupling.[11] The FM-DE (spin) and J-T distortion (orbit) are delicately balanced near $T_C/T_{IM}$. As mentioned earlier the dominance of the shape anisotropy results in an in plane easy axis and hence application of even a small magnetic field parallel to the film plane destroys this balance in favour of the FMM and the resistivity decreases sharply. Further, owing to the anisotropic nature of the enhanced J-T distortions, the magnetic field response of the out of plane and inplane components may be different. This is because the anisotropic J-T distortions could alter the spin-orbit interaction of the system via changing the magnetic interaction through the manganese ions. Thus the enhancement in the AMR could be attributed to (i) enhanced carrier localization, (ii) anisotropic nature of the J-T distortion and (iii) the dominance of the shape anisotropy.

In conclusion we have systematically studied the correlation between carrier localization or bandwidth reduction and the AMR in $Nd_{0.55-x}Sm_xSr_{0.45}MnO_3$. It has been demonstrated that J-T distortion is enhanced by the quenched disorder due to (i) the reduced average size of the rare earth site and (ii) the size disorder due to the RE/AE cations. Enhanced J-T distortion favours the growth of AFM-CO phases and results in stronger carrier localization. This in turn is observed to cause (i) a decrease in the IMT, (ii) sharper IMT, and (iii) large enhancement in low field MR and AMR. The enhancement in AMR has been explained in terms of (i) the enhanced spin fluctuations, (ii) the anisotropic nature of the J-T distortions and (iii) strong spin-orbit coupling due unquenching of the orbital angular momentum of $t_{2g}^3$ & $e_g^1$ configuration in the strong carrier localization regime (higher x).

**Acknowledgments**

MKS is thankful to CSIR, New Delhi for a senior research fellowship. Continuous support and encouragement of DNPL is thankfully acknowledged.

**Figure Captions**

Figure 1: X-ray diffraction pattern of $Nd_{0.55-x}Sm_xSr_{0.45}MnO_3$ films. Both film and substrate peaks are indexed.

Figure 2: Temperature dependent resistivity for all the $Nd_{0.55-x}Sm_xSr_{0.45}MnO_3$ films. Inset shows the variation of $T_{IM}$ with $<r_A>$ and $\sigma^2$.

Figure 3: Variation of AMR with temperature. Inset shows the variation of $AMR^{peak}$ with $<r_A>$ and $\sigma^2$.

Figure 4: Variation of peak AMR and $T_{IM}$ with activation energy ($E_A$) for all the $Nd_{0.55-x}Sm_xSr_{0.45}MnO_3$ films.

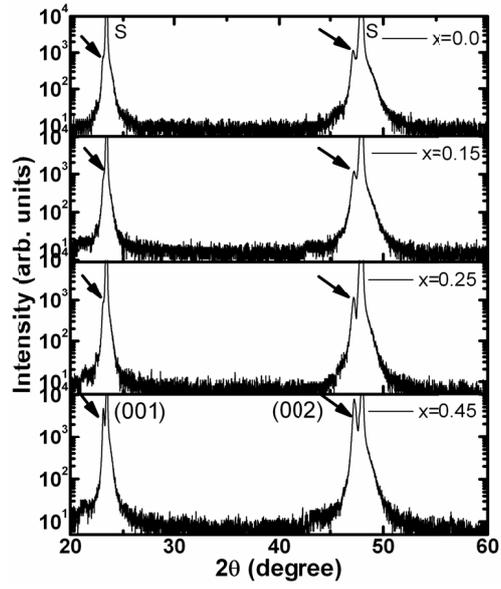

**Fig. 1**

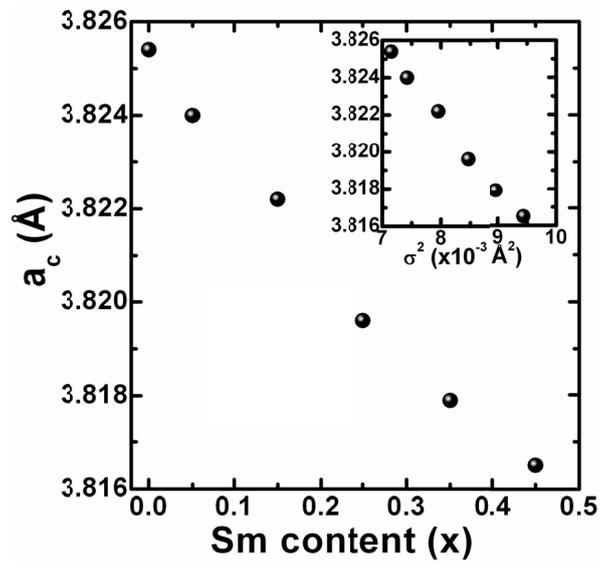

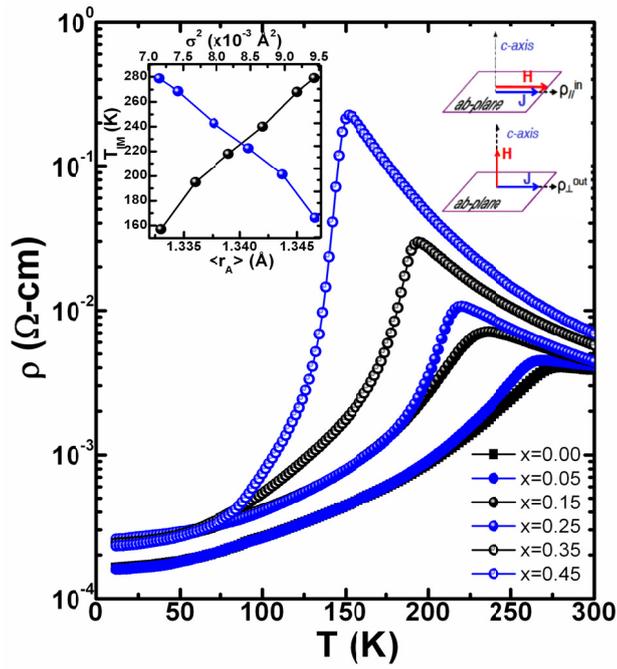

**Fig. 3**

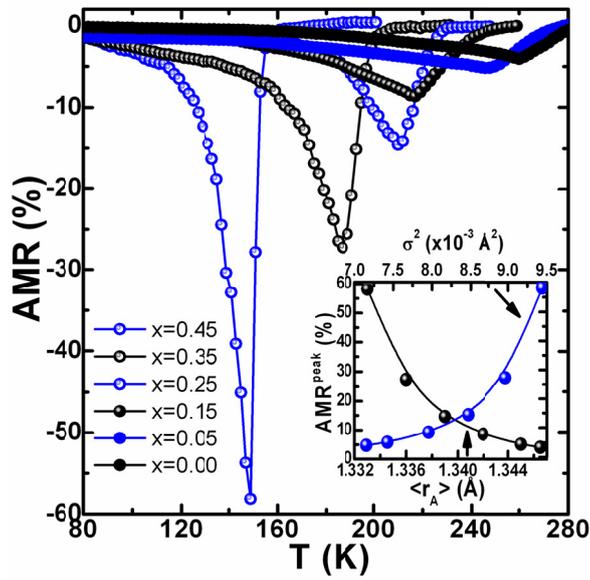

**Fig. 4**

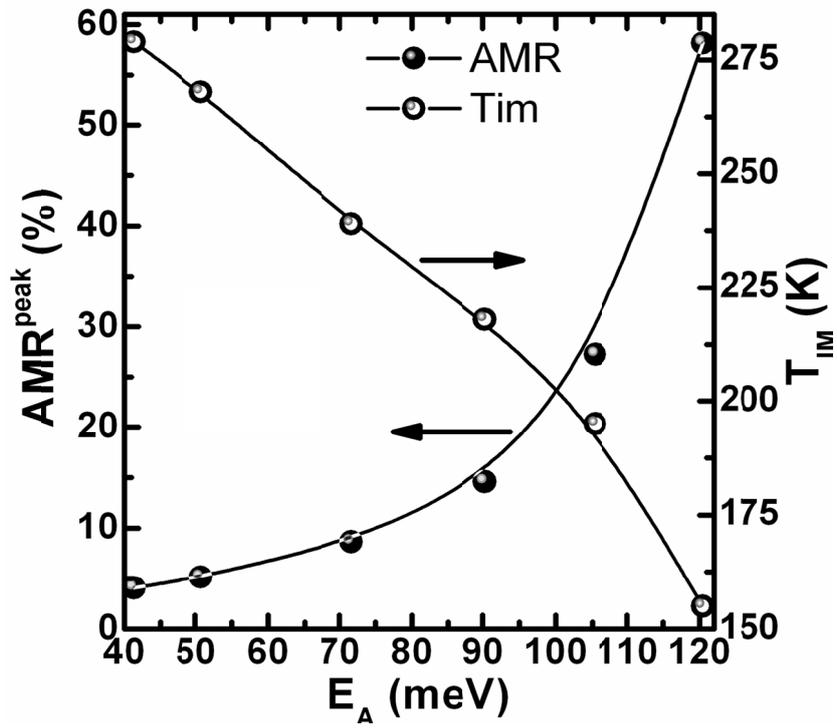

**Fig. 5**